# Accurate Enthalpies of Formation of Astromolecules: Energy, Stability and Abundance

Emmanuel E. Etim and Elangannan Arunan*

Inorganic and Physical Chemistry Department, Indian Institute of Science Bangalore,

India-560012

*email: arunan@ipc.iisc.ernet.in

**ABSTRACT:** Accurate enthalpies of formation are reported for known and potential astromolecules using high level ab initio quantum chemical calculations. A total of 130 molecules comprising of 31 isomeric groups and 24 cyanide/isocyanide pairs with atoms ranging from 3 to 12 have been considered. The results show an interesting, surprisingly not well explored, relationship between energy, stability and abundance (ESA) existing among these molecules. Among the isomeric species, isomers with lower enthalpies of formation are more easily observed in the interstellar medium compared to their counterparts with higher enthalpies of formation. Available data in literature confirm the high abundance of the most stable isomer over other isomers in the different groups considered. Potential for interstellar hydrogen bonding accounts for the few exceptions observed. Thus, in general, it suffices to say that the interstellar abundances of related species are directly proportional to their stabilities. The immediate consequences of this relationship in addressing some of the whys and wherefores among astromolecules and in predicting some possible candidates for future astronomical observations are discussed. Our comprehensive results on 130 molecules indicate that the available experimental enthalpy of formation for some molecules, such as NaCN, may be less reliable and new measurements may be needed.

# INTRODUCTION

The importance of chemistry in the interstellar space cannot be overemphasised. The interstellar space is not just an open vacuum dotted with stars, planets and other celestial formations as it is considered to be in popular perception; rather it consists of a bizarre mixture of both familiar molecules such as water, ammonia etc., and a large number of exotic ones such as radicals, acetylenic carbon chains, highly reactive cationic and anionic species, carbenes and high molecular isomers that are so unfamiliar in the terrestrial laboratory that chemists and astronomers have termed them "non-terrestrial". The development of radio-astronomical techniques and the close collaboration between laboratory spectroscopists and astrophysicists have resulted in the detection of over 200 different molecular species in the interstellar space largely via their rotational emission spectra during the last few decades. These molecules are used as probes of astrophysical phenomena. The density and temperature of the gas phase species observed in the interstellar medium (ISM) are determined from the observed spectra (rotational and vibrational) of these species. Of course these molecules are exciting clues to the chemical origin of life and serve as powerful tool in addressing the unanswered chemistry question; how the simple molecules present on the early earth may have given rise to the complex systems and processes of contemporary biology.[1-6]

Despite the importance of these molecules, not much is known about how they are formed under the low temperature and low density conditions of the interstellar clouds. This has led to the lack of a consensus on how most of these molecules; especially the complex (those with six atoms and above) ones are formed in ISM. Gas phase reactions and reactions that occur on the surface of the interstellar dust grains are the dominant processes believed to be responsible for the formation of these molecules. Molecular hydrogen plays a pivotal role in the formation of other molecular species in ISM. Among the known interstellar and circumstellar species, some of the common features observed include isomerism, successive hydrogen addition and periodic trends. These serve as pointers towards how these molecules are formed in the interstellar medium.[7,8]

Of all the observed concepts existing among interstellar molecules, the impact of isomerism appears to be very conspicuous. It is now obvious that isomerism plays a crucial role in interstellar chemistry as more and larger isomeric species are being detected in the interstellar medium. Excluding the diatomic species, a number of hydrogen saturated species and other special species like the $C_3$, $C_5$, which cannot form isomers, about 40% of all interstellar molecules have isomeric counterparts.[6,7] The clear existence of isomerism among interstellar molecules suggests that molecular formation routes for isomeric species may have common precursors for their formation processes.[7,9] With the concept of isomerism among interstellar molecules almost becoming an "established" chemistry, it can therefore be explored in unravelling other basic chemistries among the interstellar species. Astronomical searches for isomeric analogues of known interstellar species have been both successful and unsuccessful in many cases. Since interstellar isomeric species are considered to possibly have common precursors for their formation. The question arises *"why are some isomeric species observed in the interstellar space and others not?"*

In addressing this question and other whys and wherefores among the astromolecules, we employ high level quantum chemical calculations with the aim of extensively investigating the relationship; energy (enthalpy of formation), stability and abundance (ESA) among interstellar molecules which could influence the astronomical observation of some molecules at the expense of others. For this investigation, we have considered 130 molecules comprising of 31 isomeric groups (with at least one molecule astronomically observed from each isomeric group considered) and 24 cyanide/isocyanide pairs with atoms ranging from 3 to 12. To the best of our knowledge, the extensive investigation of this relationship under consideration has not been reported in the literature so far.

**COMPUTATIONAL METHODS**

The Gaussian 09 suite of programs is employed for all the quantum chemical calculations reported in this work.[10] A few of the molecules considered in this study have experimentally measured standard enthalpy of formation ($\Delta_f H^0$) while many of them do not. Theoretical methods that can predict accurate enthalpies of formation for the molecules with experimentally known $\Delta_f H^0$ values are highly desirable as such methods should by extension be able to predict enthalpies of formation of similar molecules with no experimentally measured values to chemical accuracy. The compound methods which combine both the Hartree-Fock and Post-SCF methods offer high accuracy at less computational cost. In this study, the Weizmann 1 and Weizmann 2 theory represented as W1U and W2U respectively and the Gaussian methods; G3, G4 and G4MP2 are employed in determining the standard enthalpies of formation for all the molecules considered in this study.[11,12,13] While the Weizmann methods (W1U and W2U) employ different levels of theory for geometry optimization, zero-point energy, single point calculations and energy computations, the Gaussian G4 and G4MP2 theories use the same method in their geometry optimization and zero-point energy calculations while different methods are employed in their single point calculations and energy computation. The reported zero-point corrected standard enthalpies of formation of all the molecules considered in this study were calculated from the optimized geometries of the molecules at the levels of theory mentioned above. In characterizing the stationary nature of the structures, harmonic vibrational frequency calculations were used with equilibrium species possessing all real frequencies.

**Atomization energies and enthalpy of formation**

Considering a wide range molecules with known and unknown enthalpies of formation, the total atomization energies method is more advantageous than methods like isodesmic and Benson group additivity. With a good computational method and accurate experimental values of standard enthalpy of formation of the constituents' elements involved, very high accurate enthalpies of formation can be estimated for different set molecular systems. Atomization energies (sometimes synonymously referred to as the total dissociation energies ($D_o$)) were evaluated using the calculated values of energies (sum of electronic and zero-point energy corrections) with the methods described in the computational methods above. For a reaction,

$$A_2B \rightarrow 2A + B \qquad (1)$$

The expression for computing the atomization energy of the molecule ($A_2B$) is given as;

$$\sum D_0(A_2B) = 2E_0(A) + E_0(B) - E_0(A_2B) \qquad (2)$$

In calculating the enthalpy of formation ($\Delta_f H^0$) at 0 K for all the molecules reported in this study, the experimental values of standard enthalpy of formation of elements C, H, O, N, Na, Mg, Al, Si and S reported in literature[14] were used. These values are reported in the Table 1. Values are given in kcal/mol.

The enthalpy of formation at 0 K is calculated using the following expression**:**

$$\Delta H_f^0(A_2B, 0K) = 2\Delta H_f^0(A, 0K) + \Delta H_f^0(B, 0K) - \sum D_0(A_2B) \qquad (3)$$

The enthalpy of formation at 298 K is calculated using the following expression:

$$\Delta H_f^0(A_2B, 298K) = \Delta H_f^0(A_2B, 0K) + (H_{A_2B}^O(298K) - H_{A_2B}^0(0K)) - \left[2\{H_A^0(298K) - H_A^0(0K)\} + \{H_B^0(298K) - H_B^0(0K)\}\right]$$

(4)

Thermal energies were calculated using rigid-rotor harmonic-oscillator approximations, in built in Gaussian code. The above description for calculating enthalpy of formation is only valid for neutral molecules. For ions, this is obtained from the enthalpy of formation of the corresponding neutral species and the calculated ionization potential (if it is a cation) or the electron affinity (if it is an anion).

**RESULTS AND DISCUSSION**

Experimental values of enthalpy of formation for some of the molecules considered in this study are known. These values were taken from the NIST[15] database (unless otherwise stated) and are reported in some of the Tables (as expt). Among the different high level quantum chemical calculation methods employed in this study, the G4 method gives the best estimate of the enthalpy of formation for the different molecules considered in this study. The difference between the theoretically calculated enthalpy of formation and the experimentally measured enthalpy of formation is within a few kcal/mol for molecules whose experimental enthalpies of formation are known. The calculated enthalpies of formation are also subject to the uncertainties in the experimental values of the standard enthalpy of formation of the elements used in calculating the enthalpy of formation at 0 K. All the reported enthalpies of formation (from both theory and experiment) in this study are reported in kcal/mol and at 298.15K. The different plots (Figures 1-6, 8-10, 12) reported in this paper are based on the enthalpy of formation values obtained with the G4 method. The following subsections discuss the results obtained using the various methods employed in this study. The different isomeric groups are grouped according to the number of atoms, starting from 3 to 12.

**Isomers with 3 atoms:** The zero-point corrected standard enthalpies of formation ($\Delta_f H^O$) for the different isomeric groups with three atoms considered in this study are shown in Table 2, with their current astronomical status. The G4 method estimates the enthalpy of formation for both HCN and HNC (whose experimental enthalpy of formation values are known) to chemical accuracy, i.e., the difference between the theoretically calculated values and the experimentally measured values are within ±1kcal/mol. The $CNO^-$ isomeric group is the only ion considered in this study. For the $OCN^-$, the calculated value with the W1U method is in excellent agreement with the experimental value. From Table 2, the zero-point corrected standard enthalpies of formation at the G4 and G4MP2 levels of theory are approximately the same while there are slight differences in the values obtained with the W1U and W2U methods.

In all the 6 isomeric groups considered here, the isomers with lower enthalpies of formation have all been observed in the interstellar space while only 3 of the isomers with higher enthalpies of formation have been observed. This implies that the lower the enthalpy of formation, the more stable the molecule, and the higher the stability of a molecule, the higher its abundance in the interstellar medium which makes it easy for the astronomical observation of such molecule.

In almost all the cases considered here, where both isomers have been observed, it is the most stable isomer that was first observed before the less stable one (HCN before HNC, MgNC before MgCN), implying that the most stable isomer, which is likely to be the most abundant, is astronomically easier to be detected than the less stable isomer.[16-26] HCN which is more stable than HNC is found to be more abundant than HNC in different molecular clouds[27,28]. This is also the case for the MgNC/MgCN abundance ratio measured in the asymptotic giant branch (AGM) stars[21,20]. AlNC also has a lower enthalpy of formation than AlCN and it has been observed while AlCN is yet to be observed.

Figure 1 depicts the plot of the $\Delta_f H^O$ for the molecules with 3 atoms considered in this study. It is clear from the plot that the non-observed molecules (indicated with open symbols) are the ones with higher $\Delta_f H^O$ values in their respective isomeric groups compared to the ones that have been astronomically observed (indicated with filled symbols).

**Isomers with 4 atoms:** Table 3 gives the zero-point corrected standard enthalpies of formation, $\Delta_f H^O$ in descending order of magnitude for isomers with 4 atoms considered in this study. There is a dearth of information regarding the experimental standard enthalpy of formation values for these molecules. There is a marked difference between the reported experimental $\Delta_f H^0$ value of HNCO and those predicted by all the methods considered here.

Molecules with the empirical formula CHNO are the simplest species which contain the four most important biogenic elements; carbon, hydrogen, nitrogen and oxygen. As a result, their astronomical observations are important from both the astrobiological and prebiotic chemistry perspectives.[32]

Just like in the previous case (isomers with 3 atoms), we observed the energy, stability and abundance (ESA) relationship among the isomers with 4 atoms. From the results presented in Table 3, the isomers with the lower enthalpies of formation; isocyanic acid (HNCO), cyanic acid (HOCN) and fulminic acid (HCNO) have all been observed in different sources in the interstellar space[33-36] while isofulminic(HONC) acid with higher enthalpy compared to the previous three isomers has not been observed. As HNCO is more stable than HCNO, it is more abundant than HCNO in the different molecular clouds where it has been observed.[38]

This trend is also observed for the $C_2HN$ isomeric group, where only $HC_2N$ which has the lowest enthalpy of formation has been astronomically observed.[37] Both HSCN and HNCS have been astronomically observed.[3, 39] However, it is interesting to note that the most stable isomer, HNCS, was observed long before (1979) the least stable isomer (2009). This is also the case in the CHNO isomers where the most stable isomer, HNCO, was observed in 1972, almost four decades before the other isomers. The data presented for all the four atom species, summarized in Figure 2, support the ESA relationship.

**Isomers with 5 atoms:** We present 5 different isomeric groups with 5 atoms considered in this study with their zero-point corrected standard enthalpies of formation and their current astronomical status in Table 4. The experimental enthalpy of formation of ketene is in excellent agreement with the theoretically predicted value at the G4 and G4MP2 levels. The experimental $\Delta_f H^0$ values for HCCN and $CH_2NN$ differ from the predicted value at the G4 method with 3.7 and 4.2 kcal/mol respectively.

It is crystal clear from both Table 4 and Figure 3 that there is a direct link between stability of molecules and their interstellar abundances which influences their astronomical observation. The ESA relationship is strictly followed in all the isomeric groups considered here, as only the isomers with lower enthalpies of formation in their respective groups have been astronomically observed.[42-48]

Among the $C_3HN$ isomeric group, in which more than one isomer has been observed, it is the most stable isomer, HCCCN, which was first observed (1971) before the other isomers (1992), as seen in previous cases discussed above. The $HC_3N$, the most stable isomer of the $C_3HN$ isomeric group, is also found to be more abundant than the other two known isomers: $HC_2NC$ and $HNC_3$ in all the astronomical sources where they have been observed [42,43,49,50]. HCNCC has received the attention of many investigators. However, even with the available literature about HCNCC, it has not been observed probably due to its high enthalpy of formation as compared to other isomers of $C_3NH$. This is still in agreement with the ESA relationship among interstellar molecules.

**Isomers with 6 atoms:** Both G4 and G4MP2 methods accurately predict the enthalpies of formation of methyl cyanide and methyl isocyanide in good agreement with the reported experimental values while the W2U method does better for formamide (Table 5). These accurate predictions of enthalpies of formation for molecules are good omens for the desired accuracy for the molecules with no experimental enthalpies of formation.

Of the 4 different isomeric groups (with 6 atoms) presented in Table 5 and in Figure 4, comprising of 14 molecules, 6 of them have been uniquely detected from different sources in the interstellar medium[52-57] while the remaining 8 have not been observed (except for 2H-azirine with unconfirmed astronomical observation).

The link between stability and interstellar abundance observed among isomeric species in the previous cases discussed is also noticed here. The only exception to the ESA relationship is methylene ketene which has the lowest enthalpy of formation among the $C_3H_2O$ isomers and it is yet to be astronomically observed. The ESA relationship is followed in all other cases with 6 atoms considered here. As in the previous cases, where more than one isomer has been observed, it is the most stable isomer that is first observed.

In the $C_2H_3N$ isomeric group, the most stable isomer, methyl cyanide (acetonitrile), was observed first (1971) followed by methyl isocyanide (the second most stable, 1988) and lastly ketenimine (2006). Methyl cyanide, being the most stable isomer of the group is also found to be the most abundant isomer as compared to methyl isocyanide and ketenimine in different molecular clouds and hot cores region where they have been detected.[53,55,58,83]

The recent 'detection' of 2H-azirine in the protostellar environment,[58] with only one observed transition, has been questioned.[59] Hence, it is now classified under the "non-detected" interstellar molecules. This could be linked to its low stability and probably low abundance which has resulted in the unsuccessful confirmatory searches. Propynal is about 5 times more abundant than cyclopropenone in the molecular clouds. [60, 61]

Methylene ketene, as the name suggests, is an unsaturated ketone. However, the chemistry of ketenes resembles that of carboxylic acid anhydrides which makes ketenes remarkably reactive. The high reactivity of ketenes can affect their abundances in the ISM thereby making their astronomical observations difficult. Nevertheless, ketenes remain potential candidates for astronomical observation with more sensitive astronomical instruments.

**Isomers with 7 atoms:** Table 6 lists the standard enthalpies of formation and the current astronomical status of the two isomeric groups with 7 atoms examined in this study while Figure 5 shows the plot of the $\Delta_f H^O$ for these molecules.

Fortunately, almost all (with the exception of isocyanoethene) molecules with 7 atoms considered here have experimentally measured enthalpy of formation; thus, giving ample opportunities to test the accuracy of the theoretical methods. The G4 and G4MP2 methods give excellent predictions of the zero-point corrected enthalpies of formation for the different molecules with known experimental enthalpy of formation values listed in Table 6 as compared to the W1U and W2U methods.

The trend with respect to the observed energy, stability and (interstellar) abundance relationship is nicely followed here. Unlike in other sets of isomers where some are not yet observed in the interstellar space, the 4 stable isomers of the $C_2H_4O$ family (Table 6 and Figure 5) have all been detected in the interstellar medium.[64,65,66,67] As would be expected, the most stable isomer of the $C_2H_4O$ family, acetaldehyde was first observed (1973) before the other isomers. Moreover, acetaldehyde has been observed to be present in high abundance in all the astronomical sources where it has been detected as compared to the abundances of vinyl alcohol and ethylene oxides in the same sources. [67,68,69,70]

It is important to note that the energy difference between the most stable and the least stable molecules should only be considered for species from the same set of isomers and it is not to be compared with the difference in another set of isomers. The only non-observed species here is isocyanoethene, $CH_2CHNC$, which has higher enthalpy value compared to acrylonitrile, $CH_2CHCN$ that has been observed[71].

**Isomers with 8 atoms:** Table 7 and Figure 6 give the different isomeric groups with 8 atoms considered in this study. Experimental enthalpy of formation values for the molecular species with 8 atoms from the different isomeric groups considered here (Table 7) are scarce with only acetic acid, methyl formate and cyclopropanone having experimentally measured enthalpy of formation values. The G4 and G4MP2 methods predict the enthalpy of formation of acetic acid to a very high accuracy of about 0.2 kcal/mol while that of methyl formate is 2.9kcal/mol. The experimental enthalpy of formation value for cyclopropanone is close to what is predicted by the G3 method.

The concept of isomerism among interstellar molecules is more pronounced among the interstellar complexes with eight atoms as compared to others. Of the twelve interstellar molecules[58,72-82] comprising of eight atoms, seven have isomeric (more than 50%) counterparts.

The $C_2H_4O_2$ family of isomers contains molecules of biological interest. Acetic acid is considered as a precursor for glycine; the simplest biologically important amino acid, because in the laboratory, a biomolecular synthesis of glycine occurs when acetic acid combines with amidogen cation. Glycolaldehyde is an important biomarker since it is structurally the simplest member of the monosaccharide sugars.

With the exception of methyl ketene that is yet to be astronomically observed, the data presented in Table 7 and Figure 6 show the influence of interstellar abundance on astronomical observation. The abundance of molecules in ISM has a direct correlation with their stability. The less stable molecules are of course very reactive; this high reactivity has a negative effect on their abundance in ISM as they are being transformed to other molecular species thereby making their astronomical observations difficult as compared to stable molecular species.

From Table7 and Figure 6, the data are consistent with the ESA relationship discussed in the previous cases with the exception noted above i.e. the non-observation of methyl ketene. The non-observation of methyl ketene could be traced to the same reason as noted for methylene ketene in case of isomers with 6 atoms, i.e., the carboxylic nature of ketenes which make them to be remarkably reactive as compared to propynal that has been astronomically observed.

The order of observation of the $C_2H_4O_2$ isomers follows the order of their abundance, the most abundant isomer, methyl formate, was first observed (1975) followed by the next abundant isomer, acetic acid (1997) and later by the less abundant isomer, glycolaldehyde (2000). This trend of astronomical observation has been noted in the previous cases discussed above still pointing to the fact that the isomers with lower enthalpies of formation are more stable than those with higher enthalpies of formation, the more stable isomers are more abundant in the ISM, and the isomers with higher abundances in the ISM are more easily observed astronomically as compared to those with lower abundances.

From our calculations and the experimentally measured enthalpies of formation, acetic acid is the most stable isomer of the $C_2H_4O_2$ isomeric group and as such it should also be the most abundant species in the interstellar space but this is not the case. Methyl formate is generally regarded as "interstellar weed" as a result of its high abundance in the different interstellar sources. The reason for this exception i.e. the low abundance of acetic acid as compared to methyl formate, could be due to interstellar hydrogen bonding on the surface of the interstellar dust grains which causes a greater part of acetic acid to be attached to the surface of the grains (thereby reducing its abundance) because of the presence of the acidic hydrogen (H1 in structure B of Figure 7) in the COOH group of acetic acid which is lacking in methyl formate. Of the two types of processes; gas phase chemical reactions and reactions that occur on the surfaces of dust particles, that dominate in the molecular clouds by which molecules are synthesized, the latter mechanism (reactions that occur on the surfaces of dust particles) is invoked for the formation of molecular hydrogen, $H_2$; as well as for the synthesis of larger interstellar molecules. Reactions that occur on the surfaces of dust particles create the platform for interstellar H-bonding. Interstellar hydrogen bonding is discussed in details in our fourth coming article.

It is also possible that methyl formate and propenal may be formed by more than one formation routes compared to acetic acid and methyl ketene respectively which could account for their high abundances compared to the experimentally/theoretically stable isomers. The low abundance of methyl ketene may require more sensitive instruments for its astronomical detection.

**Isomers with 9 atoms:** The enthalpies of formation and the current astronomical status of different isomeric groups with 9 atoms considered in this study are summarised in Table 8 and Figure 8. For the molecules with experimentally known enthalpies of formation, the theoretically predicted enthalpies of formation are in good agreement with results from the G3, G4 and G4MP2 methods.

The two known stable isomers of the $C_2H_6O$ isomeric groups; ethanol and dimethyl ether have been detected in the interstellar medium via their rotational transition spectra.[83,84,85] The abundance ratio of ethanol and dimethyl ether ranges from around 0.3 to 3.0 in different astronomical sources. [69,70, 86]

Among the isomeric groups with the empirical formulae $C_3H_5N$ and $C_2H_5NO$, only cyanoethane and acetamide have been observed in the interstellar medium[62,86,87] from their respective groups. The observed isomers of these sets are also the isomers with the lowest enthalpies of formation (Table 8) compared to other species in their respective groups. This observation is nicely depicted in Figure 8, with the astronomically observed molecules indicated with filled symbols while the non-astronomically observed molecules are marked with open symbols.

Without any exception, the observed the ESA relationship among interstellar isomeric species is strictly followed among the isomers with 9 atoms considered in this study.

**Isomers with 10 atoms:** Table 9 lists the isomers consisting of 10 atoms and their corresponding standard enthalpies of formation (theoretically predicted and experimentally measured, where available). Whereas the W1U and W2U methods overestimate the enthalpies of formation of these molecular species, the G4 and G4MP2 methods consistently predict the enthalpies of formation of the molecules with good agreement with the available experimental values.

In all the three different isomeric groups, the observed isomers are the isomers with the lowest enthalpies of formation compared to others in the same isomeric group.[75,89-92] In the $C_3H_6O$ isomeric group where two isomers have been astronomically observed, the most stable isomer, propanone (acetone) which is probably the most abundant isomer, was first observed (1987) before the next stable isomer, propanal, was observed (2004).

This observation is also clearly pictured in Figure 9, where the observed isomers in each group are indicated with filled symbols while the non-observed isomers are indicated with open symbols. This further supports the ESA relationship among interstellar isomeric species.

**Isomers with 11 atoms:** As for the isomers with 10 atoms, the results presented in Table 10 for the isomers with 11 atoms also point out that G4 and G4MP2 methods perform better than W1U and W2U in predicting their enthalpies of formation. Table 10 shows the results for the isomers of $C_3H_6O_2$ with their corresponding zero-point corrected enthalpies in descending order of magnitude. Figure 10 shows the current astronomical status of these molecules, with astronomically observed molecules indicated with filled symbols while the non-astronomically observed molecules are indicated with open symbols.

With the exception of propanoic acid which is the most stable isomer but yet to be astronomically observed, the astronomically observed isomers (methyl acetate and ethylformate) of this group are also the isomers with the least enthalpies of formation. [94,95]

The reason for the delayed astronomical observation of propanoic acid must be the same (interstellar H-bonding) as reason for the observed low abundance of acetic acid compared to methyl formate. In Figure 11, structures A and B show the optimized structures of propanoic acid and ethylformate respectively. H1 in structure A is the acidic H atom which can easily take part in interstellar H-bonding on the surface of the dust grains causing a greater part of propanoic acid to be attached to the surface of the interstellar dust grains, resulting in the low abundance and the subsequent difficulty in the astronomical observation of propanoic acid compared to ethylformate. An ongoing study[96] on interstellar hydrogen bonding has shown that propanoic acid is more strongly bonded to the surface of the interstellar dust as compared to other isomers of the group, thus a greater portion of it is attached to the surface of the dust grains thereby reducing its overall abundance and delaying its successful astronomical detection.

**Isomers with 12 atoms:** Despite the size of the molecule (3 to 12 atoms in this study), the G4 and G4MP2 compound models are more accurate in predicting thermochemical properties of molecules compared to the W1U and W2U methods. The zero-point corrected standard enthalpies of formation of propan-1-ol and propan-2-ol attest to this as shown in Table 11. The enthalpies of formation and the current astronomical status of different isomeric groups with 12 atoms considered in this study are summarised in Table 11 and Figure 12.

In the $C_3H_8O$ isomeric group, only ethyl methyl ether with the highest enthalpy of formation has been astronomically observed[98] while the experimentally/theoretically predicted most stable isomers, propan-1-ol and propan-2-ol are yet to be astronomically observed. The recently observed branched alkyl molecule in the ISM; isopropyl cyanide falls into the $C_4H_7N$ isomeric group. Propyl cyanide and its branched chain counterpart, isopropyl cyanide are the only astronomically observed isomers of this group.[95,99] These astronomical observations show the direct link between the stability of related molecules and their interstellar abundances, and how this link influences astronomical observations. In accordance with the ESA relationship, the observed species are also the isomers with the least enthalpies of formation in the $C_4H_7N$ isomeric group.

The delayed astronomical observations of propan-1-ol and propan-2-ol compared to ethyl methyl ether must also be due to the effect of interstellar hydrogen bonding[96]. The propensity of alcohols to form stronger hydrogen bonds than ethers is well known (Figure 11, structures C, D, and E). Ketones are generally more stable than their corresponding aldehydes because the aldehydic H atom is more acidic and so more reactive.

The low enthalpy of formation estimated for isopropyl cyanide at the G4 level of theory as compared to other isomers of the group suggests the possibility for the astronomical observation of more branched molecules in the ISM from other isomeric groups.

**Immediate Consequences of ESA Relationship**

Our knowledge of the young interdisciplinary science of astrochemistry lying at the interface of chemistry, physics, astronomy and astrophysics is still imperfect; this can be seen in the inabilities and difficulties in convincingly accounting for most of the happenings and observations in this field. The growth in the body of knowledge in this field demands bringing new ideas, insights and innovations to bear in addressing some of the challenges in this field.

The energy of a molecule is of course directly related to its stability. In this study, this stability has further been shown to be directly linked to the interstellar abundances of related molecular species (isomers in this case) thus influencing their astronomical observations. Thus, the Energy, Stability and (interstellar) Abundance (ESA) are uniquely related. But how does this ESA principle contribute to knowledge in this field in accounting for some of the observations among interstellar molecules which form a greater research area in astrochemistry and related fields; astronomy, astrobiology and astrophysics. The immediate consequences/impacts/roles of this ESA relationship in addressing some of the whys and wherefores among interstellar molecules are briefly summarised below:

**Where are Cyclic Interstellar Molecules?** With over 200 different interstellar species so far observed in the interstellar space, only 10(with the unconfirmed claimed observation of 2H-azirine) are cyclic: c-$SiC_2$, c-$C_3H$, c-$C_3H_2$, c-$H_2C_3O$, c-$C_2H_4O$, 2H-azirine, benzene, $C_{60}$, $C_{70}$ and $C_{60}^+$.[60, 67, 81,100-104] More than 10% of all the molecules considered in this study are cyclic; oxirene (Table 4), 2H-azirine, 1H-azirine, cyclopropenone (Table 5), ethylene oxide (Table 6), 1,2-dioxetane, 1,3-dioxetane, cyclopropanone, 2-cyclopropenol, 1-cyclopropenol (Table 7), cyclopropanimine, 1-azetine, 1-azabicyclo(1.1.0)butane, (Table 8), oxetane, cyclopropanol (Table 9), dioxolane, dimethyldioxirane and glycidol (Table 10). From the results, it is evident that the cyclic molecules are among the isomers with the highest enthalpies of formation and in many cases, they indeed have the highest enthalpies of formation. This explains their less stability and less abundance, subsequently resulting in the difficulty of their astronomical observation in contrast to their linear counterparts. This is a clear application of the relationship; energy, stability and abundance among interstellar molecules in addressing an important issue associated with interstellar molecules.

**Why are more Interstellar Cyanides than isocyanides?** Cyanide and isocyanide molecules account for about 20% of all the known interstellar molecules. In Tables 2-9 and 11, some of the cyanide/isocyanide pairs among interstellar molecules with their corresponding zero-point corrected enthalpies of formation and their current astronomical status are listed. In particular, the 12 astromolecules considered in this work having 3 atoms turn out to be five cyanide/isocyanide pairs and the only two anions considered in this work $ONC^-$ and $OCN^-$. In general, the cyanides have lower enthalpy of formation and they have all been observed. Clearly, the ESA relationship can explain the abundance of cyanide over isocyanide and also the exception in AlNC/AlCN (*vide infra*). As a test of this fact, of the 24 pairs of cyanide/isocyanide species considered here, the isocyanide species are only astronomically observed in 9 pairs while the cyanide species have been astronomically observed in 21 pairs. This is a direct proof of the ESA relationship existing among interstellar molecules.

**What are the possible candidates for astronomical observation?**

Only in the cases of AlNC/AlCN and MgCN/MgNC, the isocyanide has lower enthalpy of formation than the cyanide. The difference in enthalpy of formation for the AlNC/AlCN is 7.3 kcal mol$^{-1}$ according to the best theoretical estimate and 5.4 kcal mol$^{-1}$ based on experiments and AlCN is yet to be observed. Interestingly, the difference in enthalpies of formation between MgNC/MgCN is much lower (2-3 kcal mol$^{-1}$) and both have been observed. This is similar to the enthalpy difference between NaCN/NaNC and we predict that NaNC may be observed in the near future. Curiously, the largest difference between the calculated and experimental enthalpy of formation happens to be for NaCN. We suspect that the experimental value may have some error.

With respect to larger isomeric species, possible candidates for astronomical observations can easily be predicted following the ESA relationship discussed in this work. Among others, the five molecules; methylene ketene, methyl ketene, propanoic acid, propanol and propan-2-ol which have the lowest enthalpy of formation values in their respective isomeric group but are yet to be astronomically observed. These are potential candidates for astronomical observation. From the recent observation of the branched alkyl molecule in the ISM and our calculations which showed the branched molecule to be the most stable molecule among its isomers paints a picture of further astronomical observation of other branched molecules which could also be more stable among its isomers.

**Conclusions:** Among the different compound models employed in estimating accurate enthalpies of formation for known and potential astromolecules, the Gaussian G4 and G4MP2 methods have proven to be consistently good in predicting accurate enthalpies of formation that are in good agreement with the available experimental values. From the results, the ESA relationship: energy, stability and (interstellar) abundance, is found to exist among interstellar molecules. From this relationship, interstellar abundances of related species are directly proportional to their stabilities. This influences the astronomical observations of some related molecular species at the expense of others.

Some of the immediate consequences of this relationship in addressing some of the whys and wherefores among interstellar molecules such as "Where are cyclic interstellar molecules? Why are more interstellar cyanides than isocyanides? What are the possible candidates for astronomical observation?" etc, have been highlighted in this article. The few exceptions are well rationalized on the grounds of interstellar hydrogen bonding and sensitivity of astronomical instruments. It is hoped to be a useful tool in the fields of astrochemistry, astronomy, astrophysics and other related disciplines.

**Acknowledgement**: Authors acknowledge Dr. Sai G. Ramesh for the computational facility at the Inorganic and Physical Chemistry Department and EEE acknowledges a research fellowship from the Indian Institute of Science, Bangalore.

**References**

1. P. Thaddeus, M. Guélin, R. A. Linke, *ApJ.*, 1971, **246**, 41-45.
2. E. Herbst, *Chem. Soc. Rev.*, 2001, **30**, 168-176.
3. E. E. Etim, and E. Arunan, E. *Planex Newsletter*, 2015, **5**(2):16-21.
4. T. D. Crawford, J. F. Stanton, J. C. Saeh, H. F. Schaefer, *J. Am. Chem. Soc*. 1999, **121**, 1902.
5. G. Wlodarczak, *J. Mol. Structure*, 1995, **347**, 131.

6. A. J. Remijan, J. M. Hollis, F. J. Lovas, D. F. Plusquellic, P. R. Jewell, *ApJ*, 2005, **632**, 333-339.
7. J. M. Hollis, In Proc. *IAU Symposium No*. **231**, 2005. D. C. Lis; G. A. Blake; E. Herbst, eds.
8. A. G. G. M. Tielens, *Rev. Mod. Phys*, 2013, **85**:1021-1081.
9. R. L. Hudson, M. H. Moore, *Icarus*, 2004, **172**, 466-478.
10. M. J., Frisch, G. W. Trucks, H. B. Schlegel, G. E. Scuseria, M. A. Robb, J. R. Cheeseman, G. Scalmani, V. Barone, B. Mennucci, G. A. Petersson, Gaussian 09, revision D.01;Gaussian, Inc., Wallingford, CT, 2009.
11. J. M. L. Martin, G. de Oliveira, *Jour. Chem. Phys.*, 1999, **111**, 1843-1856.
12. L A. Curtiss, P. C. Redfern, K. Raghavachari, *Jour. Chem. Phys*, 2007, **126**, 084108-12.
13. L. A. Curtiss, P. C. Redfern, K. Raghavachari, *Jour. Chem. Phys*, 2007, **127**, 124105-8.
14. L. A. Curtiss, K. Raghavachari, P. C. Redfern, J. A. Pople, J. A. Jour. Chem. Phys, 1997, **106**, 1063-1079.
15. http://webbook.nist.gov/chemistry/   Accessed in May 2016
16. B. Zuckerman, M. Morris, P. Palmer, B. E. Turner, *ApJ*, 1972, **173**,125-129.
17. G. L. Blackman, R. D. Brown, P. D. Godfrey, H. I. Gunn, *Nature*, 1976, **261**:395.
18. B. E. Turner, T. C. Steimle, L. Meerts, *ApJ*, 1994, **426**,97-100.
19. L. M. Ziurys, A. J. Apponi, M. Guélin, J. Cernicharo, *ApJ*, 1995, **445**,47-50.
20. K. Kawaguchi, E. Kagi, T. Hirano, S. Takano, S. Saito, *ApJ*, 1993, **406**, 39-42.
21. M. Guélin, J. Cernicharo, C. Kahane, J. Gomez-Gonzales, *A&A*, 1986, **157**, L17-L20.
22. M. Guelin, R. Lucas, J. Cernicharo, *A&A,* 1993, **280**,19-22.
23. M. Guélin, S. Muller, J. Cernicharo, A. J. Apponi, M. C. McCarthy, C. A. Gottlieb, P. Thaddeus, *A&A*, 2000, **369**, L 9-L12.
24. M. Guélin, S. Muller, J. Cernicharo, M. C. McCarthy, P. Thaddeus, *A&A*, 2004, **426**, L49-L52.
25. L. M. Ziurys, C. Savage, J. L. Highberger, A. J. Apponi, M. Guélin, J. Cernicharo, *ApJ*, 2002, **564**, 45-48.
26. B. T. Soifer, R. C. Puetter, R. W. Russell, S. P. Willner, P. M. Harvey, F. C. Gillett, *ApJ*, 1979, **232**, 53-57.
27. P. P. Tennekes, J. Harju, M. Juvela, L. V. Tóth, *A&A*, 2006,**456**, 1037-1043.
28. W. M. Irvine, F. P. Schloerb, *ApJ*, 1984, **282**, 516-521.
29. N. J. DeYonker, D. S. Ho, A. K. Wilson, T. R. Cundari, T. R. *J. Phys. Chem. A*. 2007, **111**, 10776-10780.

30. G. Meloni and K. A. Gingerich, *J. Chem. Phys.,* 1999, **111**, 969-972.
31. S. E. Bradorth, E. H. Kim, D. W. Arnold, D. M. Neumark, *J. Chem. Phys.*, 1993, **98**, 800-810.
32. D. Poppinger, L. Radom, J. A. Pople, *J. Amer. Chem. Soc.,* 1977, **99**, (24), 7806-7816.
33. L. E. Snyder, D. Buhl, *ApJ.*, 1972, **177**, 619-623.
34. S. Brunken, C. A. Gottlieb, M. C. McCarthy, P. Thaddeus, *ApJ* 2009, **697**, 880-885.
35. S. Brunken, A. Belloche, S. Martin, L. Verheyen, K. M. Menten, *A&A*, 2010, **516**, A109
36. N. Marcelino, J. Cernicharo, B. Tercero, E. Roueff, *ApJ.*, 2009, **690**, 27-30.
37. M. Guélin, J. Cernicharo, *A&A*, 1991, **244**, L21-L24.
38. D. T. Halfen, L. M. Ziurys, S. Brünken, C. A. Gottlieb, M. C. McCarthy, P. Thaddeus, *ApJ*, 2009, **702**, 124-127
39. M. A. Frerking, R. A. Linke, *ApJ*, 1979, **234**, 143-145.
40. J. C. Poutsma, S. D. Upshaw, R. R. Squires, , & P. G. Wenthold, *J. Phys. Chem. A*, 2002, **106**, 1067-1073.
41. http://chemister.ru/Database/properties-en.php?dbid=1&id=3123 Accessed in May 2016.
42. K. Kawaguchi, M. Ohishi, S. I. Ishikawa, N. Kaifu, *ApJ*, 1992, **386**, 51-53.
43. K. Kawaguchi, S. Takano, M. Ohishi, S. I. Ishikawa, K. Miyazawa, N. Kaifu, K. Yamashita, S. Yamamoto, S. Saito, S. Ohshima, Y. Endo, *ApJ*, 1992, **396**, 49-51.
44. A. J. Remijan, J. M. Hollis, F. J. Lovas, W. D. Stork, P. R. Jewell, D. S. Meier, *ApJ*, 2008, **675**, L85-L88.
45. B. E. Turner, H. S. Liszt, N. Kaifu, A. G. Kisliakov, ApJ, 1975, **201**, 149.
46. B. E. Turner, *ApJ*, 1971, **163**, 35-39.
47. W. M. Irvine, P. Friberg, A. Hjalmarson, S. Ishikawa, N. Kaifu, K.Kawaguchi, S. C. Madden, H. E. Matthews, M. Ohishi, S. Saito, H.Suzuki, *ApJ*, 1988, **334**, 107-111.
48. W. M. Irvine, R. D. Brown, D. M. Craig, P. Friberg, P. D. Godfrey, N. Kaifu, H. E. Matthews, M. Ohishi, H. Suzuki, H. Takeo, *ApJ*, 1988, **335**,89-93.
49. J. R. Pardo, J. Cernicharo, J. R. Goicoechea, M. Guélin, A. Ramos, *ApJ*, **661**, 250-261.
50. P. D. Gensheimer, *Ap&SS*, 1997, **251**, 199-202.
51. M. Sablier, & T. Fujii, *Chemical Reviews*, 2002, **102**,2855-2924.
52. P. M. Solomon, K. B. Jefferts, A. A. Penzias, R. W. Wilson, *ApJ.*, 1971, **168**, 107-110.
53. J. Cernicharo, C. Kahane, M. Guélin, J. Gomez-Gonzalez, *A&A*, 1988, **189**, 1-2.
54. J. Cernicharo, M. Guélin, J. R. Pardo, *ApJ*, 2004, **615**, 145-148.
55. F. J. Lovas, J. M. Hollis, A. J. Remijan, P. R. Jewell, *ApJ*, 2006, **645**,137-140.

56. F. J. Lovas, A. J. Remijan, J.M. Hollis, P. R. Jewell, L. E. Snyder, *ApJ.*, 2006, **637**, 37-40.
57. R. H. Rubin, G. W. Swenson, R. C. Solomon, Jr, H. L. Flygare, *ApJ*, 1971, **169**, 39-44.
58. S. B. P. Charnley, P. Ehrenfreund, Y. J. Kuan, *AcSpA*, 2001, **57**, 685-704.
59. http://www.astrochymist.org/astrochymist_nondetections.html Accessed May 2016.
60. B. E. Turner, *ApJS*, 1991, **76**, 617-686.
61. J. M. Hollis, F. J. Lovas, A. J. Remijan, P. R.Jewell, V. V. Ilyushin, I. Kleiner, *ApJ.*, 2006, **643**, 25-28.
62. V. M. Emel'yanenko, S. P. Verevkin, M. A. Varfolomeev, *J. Chem. Eng. Data*, 2011, **56**, 4183–4187.
63. https://www.chemeo.com/cid/27-566-4/Propynal,%20dimethylacetal.pdf Accessed May 2016
64. N. Fourikis, M. W. Sinclair, B. J. Robinson, P. D. Godfrey, R. D. Brown, AuJPh, 1974, **27**, 425-430.
65. W. Gilmore, M .Morris, P. Palmer, D. R. Johnson, F. J. Lovas, B. E. Turner, *ApJ*, 1976, **204**,43.
66. B. E. Turner, A. J. Apponi, *ApJ*, 2001, **561**, 207-210.
67. J. E. Dickens, W. M. Irvine, M. Ohishi, M. Ikeda, S. Ishikawa, A. Nummelin, A. Hjalmarson, *ApJ*, 1997, **489**,753-757.
68. A.Nummelin, J.Dickens, P. Bergman, A. Hjalmarson, W. M. Irvine, M. Ikeda, M. Ohishi, *ApJS*, 1998, **117**:427-529.
69. A. Nummelin, P. Bergman, A. Hjalmarson, P. Friberg, W. M. Irvine, T. J. Millar, M. Ohishi, S. Saito, *A&A*, 1998, **337**, 275-286.
70. M. Ikeda, M. Ohishi, J. E. Dickens, P. Bergman, A. Hjalmarson, W. M. Irvine, *ApJ*, 2001, **560**, 792-805.
71. F. F. Gardner, & G. Winnewisser, G. *ApJ*, 1975, **195**, L127-L130.
72. M. Guélin, J. Cernicharo, M. J. Travers, M. C. McCarthy, C.A. Gottlieb, P. Thaddeus, M. Ohishi, S. Saito, S. Yamamoto, *A&A*, 1997, **317**, L1-L4.
73. J. M.Hollis, F. J. Lovas, P. R. Jewell, *ApJ*, 2000, **540**, 107-110.
74. J. M. Hollis, P. R. Jewell, F. J. Lovas, A. Remijan, H. Møllendal, *ApJ*, 2004, **610**, 21-24.
75. N. W. Broten, J. M. MacLeod, L. W. Avery, W. M. Irvine, B. Höglund, P. Friberg, P. Hjalmarson, *ApJ*, 1984, **276**, 25-29.
76. W. D. Langer, T. Velusamy, T. B. H. Kuiper, R. Peng, M. C. McCarthy, M. J. Travers, A. Kovács, C. A. Gottlieb, P. Thaddeus, *ApJ*, 1997, **480**, 63-66.

77. E. Churchwell, G. Winnewisser, *A&A.*, 1975, **45**,229-231.
78. R. D. Brown, J. G. Crofts, P. D. Godfrey, F. F. Gardner, B. J. Robinson, J. B. Whiteoak, *ApJ*, 1975, **197**:L29-L31.
79. A. Belloche, K. M. C. Menten, C. Comito, H. S. P. Müller, P.Schilke, J. Ott, S. Thorwirth, C. Hieret, *A&A.*, 2008, **482**,179-196.
80. R. A. Loomis, D. P. Zaleski, A. L. Steber, J. L. Neill, M. T. Muckle, B. J. Harris, J. M. Hollis, P. R. Jewell, V. Lattanzi, F. J. Lovas, O. Martinez, M. C. McCarthy, A. J. Remijan, B. H. Pate, J. F. Corby, *ApJL*, 2013, **765**, L9.
81. J. Cernicharo, A. M. Heras, A. G. G.M. Tielens, P. R. Pardo, F. Herpin, M. Guélin, L. B. F. M. Waters, *ApJ*, 2001, **546**, L123-L126.
82. A. J. Remijan, L. E. Snyder, B. A. McGuire, H. L Kuo, L. W. Looney, D. N. Friedel, G. Y. Golubiatnikov, F. J. Lovas, V. V. Ilyushin, E. A. Alekseev, S. F. Dyubko, B. J. McCall, J. M. Hollis, *ApJ.*, 2014,**783**:77.
83. B. Zuckerman, B. E. Turner, D. R. Johnson, F. O. Clark, F. J. Lovas, N. Fourikis, P. Palmer, M. Morris, A. E. Lilley, J. A. Ball, C. A. Gottlieb, H. Penfield, *ApJ.*, 1975, **196**, 99-102.
84. J. C. Pearson, K. V. L. N. Sastry, E. Herbst, F. C. De Lucia, *ApJ*, 1997, **480**, 420-431.
85. L. E. Snyder, D. Buhl, P. R. Schwartz, F. O. Clark, D. R. Johnson, F. J. Lovas, P. T. Giguere, *ApJ.*, 1974, **191**, 79-82.
86. G. J. White, M. Araki, J. S. Greaves, M. Ohishi, N. S. Higginbottom, *A&A*, 2003, **407**: 589-607.
87. D. R. Johnson, F. J. Lovas, C. A. Gottlieb, E. W. Gottlieb, M. M. Litvak, M.; Guélin, P. Thaddeus, *ApJ.*, 1977, **218**, 370-376.
88. J. Cioslowski, M. Schimeczek, G. Liu, V. Stoyanov, *J. Chem. Phys.*, 2000, **113**, 9377-9388.
89. F. Combes, M. Gerin, A. Wooten, G. Wlodarczak, F. Clausset, P. I. Encrenaz, *A&A.*, 1987, **180**,L13-L16.
90. J. M. Hollis, F. J. Lovas, P. R. Jewell, L. H. Coudert, *ApJ*, 2002, **571**, 59-62.
91. L. E. Snyder, F. J. Lovas, D. M. Mehnringer, N. Y. Miao, Y. -J. Kuan. J. M. Hollis, P. R. Jewell, *ApJ*, 2002,**578**, 245-255.
92. L. E. Snyder, J. M. Hollis, P. R. Jewell, F. J. Lovas, A. Remijan, *ApJ*, 2006, **647**,412-417.
93. https://pqr.pitt.edu/mol/YOXHCYXIAVIFCZ-UHFFFAOYSA-N# Accessed May 2016.
94. A. Belloche, R. T. Garrod, H. S. P. Müller, K. M. Menten, C. Comito, P. Schilke, *A&A.*, 2009, **499**, 215-232.

95. B. Tercero, I. Kleiner, J. Cernicharo, H. V. L. Nguyen, A. López, G. M. Muñoz Caro, *ApJL.*,2013, **770**, 13
96. E. E. Etim, E. Arunan, Manuscript to be submitted.
97. H. K. Hall, Jr, & J. H. Baldt, *J. Amer. Chem. Soc*, 1971, **93**, 140-145.
98. G. W. Fuchs, U. Fuchs, T. F. Giesen, F. Wyrowski, *A&A*, 2005, **444**, 521-530.
99. A. Belloche, R. T. Garrod, H. S. P.Müller, K. M. Menten, *Science*, 2014, **345**, 1584-1587.
100. J. M. Hollis, A. J. Remijan, P. R. Jewell, F. J. Lovas, *ApJ*, 2006, **642**, 933-939.
101. P. Thaddeus, S. E. Cummins, R. A. Linke, *ApJ*, 1984, **283**, 45-48.
102. P. Thaddeus, J. M. Vrtilek, C. A. Gottlieb, *ApJ*, 1985, **299**, 63-66.
103. J. Cami, J. Bernard-Salas, E. Peeters, S. E. Malek, *Science*, 2010, **329**:1180-1192.
104. O. Berné, G. Mulas, and C. Joblin, *A&A*, 2013, **550**:L4.

Tables and Figures

**Tables**

Table 1. Experimental $\Delta_f H^0$ (0K), of elements and $H^0$ (298K) – $H^0$ (0K)

| Element | $\Delta_f H^0$ (0K) | $H^0$ (298K)-$H^0$ (0K) |
|---------|---------------------|--------------------------|
| H  | 51.63±0.01  | 1.01 |
| C  | 169.98±0.1  | 0.25 |
| O  | 58.99±0.02  | 1.04 |
| N  | 112.53±0.02 | 1.04 |
| Na | 25.69±0.17  | 1.54 |
| Mg | 34.87±0.2   | 1.19 |
| Al | 78.23±1.0   | 1.08 |
| Si | 106.6±1.9   | 0.76 |
| S  | 65.66±0.06  | 1.05 |

Table 2: $\Delta_f H^O$ for isomers with 3 atoms and current astronomical status

| Molecule | Enthalpy of formation | | | | | | Astronomical status |
|---|---|---|---|---|---|---|---|
| | W1U | W2U | G3 | G4MP2 | G4 | Expt | |
| HNC | 45.03 | 44.7 | 46.0 | 45.6 | 45.6 | 46.5±2.2 | Yes |
| HCN | 31.3 | 31.4 | 31.3 | 32.2 | 32.2 | 32.30 | Yes |
| NaNC | 36.4 | 45.3 | 28.4 | 34.5 | 37.0 | | No |
| NaCN | 34.3 | 43.8 | 28.4 | 34.5 | 34.5 | 22.5±0.5[29] | Yes |
| MgCN | 67.7 | 67.3 | 66.5 | 69.0 | 69.0 | | Yes |
| MgNC | 65.6 | 65.6 | 65.7 | 69.7 | 65.8 | | Yes |
| AlCN | 71.8 | 72.3 | 68.4 | 73.1 | 73.1 | 71.9±3.3[30] | No |
| AlNC | 65.2 | 65.2 | 63.2 | 65.8 | 65.8 | 66.5±3.3[30] | Yes |
| SiCN | 105.1 | 105.2 | 105.8 | 105.9 | 105.9 | | Yes |
| SiNC | 105.3 | 105.3 | 107.5 | 105.7 | 105.7 | | Yes |
| ONC- | 14.3 | 14.2 | 14.3 | 12.4 | 12.4 | | No |
| OCN- | -53.7 | -54.0 | -53.4 | -54.5 | -54.5 | -52.8[31] | Yes |

Table 3: $\Delta_fH^O$ for isomers with 4 atoms and current astronomical status

| Molecule | Enthalpy of formation | | | | | | Astronomical status |
|---|---|---|---|---|---|---|---|
| | W1U | W2U | G3 | G4MP2 | G4 | Expt | |
| Isofulminic acid | 55.1 | 55.5 | 55.7 | 52.8 | 52.8 | | No |
| Fulminic acid | 36.3 | 36.3 | 40.0 | 34.1 | 34.1 | | Yes |
| Cyanic acid | -3.0 | -2.2 | -3.5 | -4.4 | -4.4 | | Yes |
| Isocyanic acid | -31.6 | -31.6 | -28.8 | -33.4 | -33.4 | -23±3.1 | Yes |
| HCNC | 135.9 | 136.0 | 139.4 | 133.5 | 133.5 | | No |
| HCCN | 123.9 | 123.9 | 126.1 | 122.6 | 122.6 | 126±3.0[40] | Yes |
| HSCN | 39.7 | 39.7 | 38.3 | 38.3 | 38.3 | | Yes |
| HNCS | 28.2 | 28.8 | 29.8 | 27.1 | 27.1 | 25.0[41] | Yes |

Table 4: $\Delta_f H^O$ for isomers with 5 atoms and current astronomical status

| Molecule | Enthalpy of formation | | | | | | Astronomical status |
|---|---|---|---|---|---|---|---|
| | W1U | W2U | G3 | G4MP2 | G4 | Expt | |
| HCNCC | 163.0 | 162.2 | 167.3 | 160.9 | 160.9 | | No |
| CC(H)CN | 140.2 | 140.2 | 139.5 | 138.8 | 138.8 | | No |
| HNCCC | 135.4 | 135.4 | 140.1 | 134.0 | 133.7 | | Yes |
| HCCNC | 115.0 | 115.0 | 115.7 | 112.3 | 112.4 | | Yes |
| HCCCN | 89.7 | 89.8 | 88.8 | 88.4 | 88.3 | 84.6 | Yes |
| | | | | | | | |
| HCONC | 25.0 | 25.1 | 24.8 | 21.4 | 21.4 | | No |
| HCOCN | 13.4 | 13.4 | 11.7 | 10.6 | 10.6 | | Yes |
| | | | | | | | |
| Oxirene | 68.6 | 68.6 | 65.0 | 66.3 | 66.3 | | No |
| Ethynol | 24.8 | 24.8 | 22.2 | 23.2 | 23.2 | | No |
| Ketene | -12.5 | -12.5 | -12.1 | -15.6 | -15.6 | -14.78 | Yes |
| | | | | | | | |
| CH$_2$NC | 81.6 | 81.7 | 85.7 | 79.1 | 79.1 | | No |
| CH$_2$CN | 58.5 | 58.6 | 61.8 | 56.8 | 56.8 | 58±3[51] | Yes |
| | | | | | | | |
| NH$_2$NC | 73.6 | 73.6 | 78.2 | 72.7 | 72.7 | | No |
| CH$_2$NN | 56.9 | 56.9 | 64.5 | 55.6 | 55.6 | 51.4 | No |
| NH$_2$CN | 29.3 | 29.3 | 33.0 | 29.2 | 29.2 | | Yes |

Table 5: $\Delta_f H^O$ for isomers with 6 atoms and current astronomical status

| Molecule | Enthalpy of formation | | | | | | Astronomical status |
|---|---|---|---|---|---|---|---|
| | W1U | W2U | G3 | G4MP2 | G4 | Expt | |
| 1H-azirine | 99.4 | 99.4 | 99.1 | 96.7 | 96.7 | | No |
| 2H-azirine | 66.1 | 66.1 | 66.0 | 62.4 | 62.4 | | No |
| Ethyneamine | 58.4 | 58.4 | 59.9 | 58.1 | 58.1 | | No |
| Ketenimine | 40.5 | 40.6 | 44.7 | 38.9 | 38.9 | | Yes |
| Methyl isocyanide | 41.9 | 42.0 | 42.2 | 38.8 | 38.8 | 39±2 | Yes |
| Methyl cyanide | 18.2 | 18.2 | 17.8 | 15.8 | 15.8 | 15.74 | Yes |
| HC$_3$NC | 185.9 | 185.9 | 189.7 | 181.9 | 181.1 | | No |
| HC$_4$N | 166.8 | 166.8 | 171.0 | 163.6 | 163.6 | | Yes |
| Cyclopropenone | 40.0 | 40.0 | 43.2 | 38.1 | 38.1 | | Yes |
| Propynal | 35.2 | 35.2 | 32.1 | 31.3 | 31.3 | 29.7[62] | Yes |
| **Methylene ketene** | **29.5** | **29.5** | **30.1** | **30.8** | **30.8** | | ***No*** |
| Nitrosomethane | 16.7 | 16.8 | 17.7 | 13.5 | 13.5 | | No |
| Hydroxymethylimine | -34.0 | -33.9 | -33.6 | -34.7 | -34.7 | | No |
| Formamide | -46.7 | -46.6 | -45.0 | -47.3 | -47.3 | -44.5 (45.1±0.1[63]) | Yes |

Table 6: $\Delta_f H^O$ for isomers with 7 atoms and current astronomical status

| Molecule | Enthalpy of formation | | | | | | Astronomical status |
|---|---|---|---|---|---|---|---|
| | W1U | W2U | G3 | G4MP2 | G4 | Expt | |
| Ethylene oxide | -8.7 | -8.7 | -5.8 | -14.6 | -14.6 | -12.58 | Yes |
| Vinyl alcohol (anti) | -25.7 | -25.7 | -28.5 | -28.5 | -28.5 | -29.9±2.0 | Yes |
| Vinyl alcohol (syn) | -27.1 | -27.1 | -29.5 | -30.2 | -30.2 | -30.6 | Yes |
| Acetaldehyde | -37.5 | -37.5 | -39.6 | -42.4 | -42.4 | -40.8±0.4 | Yes |
| Isocyanoethene | 66.9 | 66.2 | 66.7 | 63.2 | 63.2 | | No |
| Acrylonitrile | 46.1 | 46.1 | 44.8 | 43.3 | 43.3 | 42.9 | Yes |

Table 7: $\Delta_f H^O$ for isomers with 8 atoms and current astronomical status

| Molecule | Enthalpy of formation | | | | | | Astronomical status |
|---|---|---|---|---|---|---|---|
| | W1U | W2U | G3 | G4MP2 | G4 | Expt | |
| CH$_2$CCHNC | 103.5 | 103.5 | 104.0 | 98.8 | 98.8 | | No |
| CH$_3$CCNC | 104.4 | 104.5 | 108.3 | 98.6 | 98.6 | | No |
| HCCCH$_2$CN | 88.8 | 88.8 | 89.1 | 86.1 | 86.1 | | No |
| CH$_2$CCHCN | 80.0 | 80.0 | 79.6 | 76.2 | 76.2 | | Yes |
| CH$_3$CCCN | 77.6 | 77.6 | 85.8 | 73.1 | 73.1 | | Yes |
| | | | | | | | |
| H$_2$NCH$_2$NC | 51.7 | 50.7 | 51.5 | 47.1 | 47.1 | | No |
| H$_2$NCH$_2$CN | 30.9 | 30.9 | 30.9 | 28.1 | 28.1 | | Yes |
| | | | | | | | |
| 1,2-dioxetane | 7.7 | 7.7 | 1.9 | -0.7 | -0.7 | | No |
| 1,3-dioxetane | -42.3 | -42.3 | -48.6 | -50.9 | -50.9 | | No |
| Glycolaldehyde | -65.3 | -65.2 | -86.0 | -70.5 | -70.5 | | Yes |
| Methylformate | -82.2 | -82.2 | -99.6 | -89.4 | -89.4 | -86.5 | Yes |
| Acetic acid | -98.0 | -98.6 | -95.9 | -103.7 | -103.7 | -103.5±0.7 | Yes |
| | | | | | | | |
| Epoxypropene | 52.7 | 52.7 | 52.4 | 47.7 | 47.7 | | No |
| 2-cyclopropenol | 32.0 | 32.0 | 26.2 | 27.4 | 27.4 | | No |
| 1-cyclopropenol | 31.9 | 31.9 | 27.3 | 25.9 | 25.9 | | No |
| Methoxyethyne | 29.8 | 29.8 | 26.4 | 23.9 | 23.9 | | No |
| Propargyl alcohol | 20.0 | 20.0 | 13.8 | 17.2 | 17.2 | | No |
| Propynol | 17.6 | 17.1 | 13.5 | 12.7 | 12.7 | | No |
| Cyclopropanone | 7.8 | 7.9 | 4.7 | 0.7 | 0.7 | 3.8±1.0 | No |
| Propenal | -10.5 | -10.5 | -13.8 | -15.8 | -15.8 | | Yes |
| *Methyl ketene* | *-14.0* | *-13.9* | ***-15.7*** | *-20.4* | *-18.1* | | *No* |

Table 8: $\Delta_f H^O$ for isomers with 9 atoms and current astronomical status.

| Molecule | Enthalpy of formation | | | | | | Astronomical status |
|---|---|---|---|---|---|---|---|
| | W1U | W2U | G3 | G4MP2 | G4 | Expt | |
| Dimethyl ether | -41.4 | -41.4 | -44.4 | -45.1 | -45.1 | -44.0±0.1 | Yes |
| Ethanol | -51.3 | -51.3 | -56.3 | -56.7 | -56.7 | -56.2±0.1[88] | Yes |
| Cyanoethoxy-amide | 77.7 | 77.7 | 76.3 | 72.5 | 72.5 | | No |
| 1-aziridnol | 22.3 | 22.4 | 19.6 | 16.0 | 16.0 | | No |
| Nitrosoethane | 10.5 | 10.5 | 9.0 | 2.8 | 2.8 | | No |
| N-methylformate | -45.0 | -46.0 | -45.7 | -52.2 | -52.2 | | No |
| Acetamide | -57.0 | -57.5 | -57.9 | -61.9 | -61.9 | -56.96±0.19 | Yes |
| 1-azabicyclo (1.1.0)butane | 72.3 | 72.331 | 67.5 | 64.4 | 64.4 | | No |
| Propargylamine | 59.2 | 59.2 | 56.6 | 56.0 | 56.1 | | No |
| Methylazaridine | 60.6 | 60.9 | 60.8 | 54.8 | 54.8 | | No |
| Cyclopropan-imine | 54.3 | 54.3 | 52.3 | 48.2 | 48.2 | | No |
| 1-Azetine | 50.4 | 50.5 | 49.9 | 43.2 | 43.2 | | No |
| N-methylene ethenamine | 42.0 | 42.1 | 40.3 | 36.9 | 36.9 | | No |
| 2-propen-1-imine | 37.0 | 37.1 | 36.5 | 36.2 | 36.2 | | No |
| Propylenimine | 37.9 | 38.0 | 37.2 | 33.0 | 33.0 | | No |
| Isocyanoethane | 37.8 | 37.9 | 32.8 | 31.8 | 31.8 | 31.7 | No |
| Cyanoethane | 16.3 | 16.3 | 13.2 | 11.0 | 11.0 | 12.30 | Yes |

Table 9: $\Delta_f H^O$ for isomers with 10 atoms and current astronomical status

| Molecule | Enthalpy of formation | | | | | | Astronomical status |
|---|---|---|---|---|---|---|---|
| | W1U | W2U | G3 | G4MP2 | G4 | Expt | |
| Dimethylperoxide | -27.0 | -27.6 | -29.2 | -35.9 | -35.9 | -30.1 | No |
| Ethylhydroperoxide | -34.0 | -33.9 | -38.3 | -41.7 | -41.7 | | No |
| Ethylene glycol | -82.2 | -82.2 | -90.9 | -87.5 | -87.5 | -92.7±0.5 | Yes |
| Oxetane | -12.4 | -12.4 | -19.1 | -21.8 | -21.8 | -19.25±0.15 | No |
| Cyclopropanol | 17.0 | -17.0 | -24.0 | -21.8 | -23.8 | -28.6±4.0[93] | No |
| 1,2-epoxypropane | -16.3 | -16.3 | -22.7 | -25.0 | -25.0 | 22.6±0.2 | No |
| 2-propene-1-ol | -22.8 | -22.8 | -29.2 | -27.9 | -27.9 | -29.5.±0.4 | No |
| Methoxyethene | -21.9 | -21.6 | -26.0 | -29.4 | -29.4 | | No |
| 1-propen-1-ol | -30.4 | -30.3 | -35.4 | -36.4 | -36.4 | | No |
| 1-propen-2-ol | -35.1 | -35.1 | -40.5 | -41.2 | -41.2 | -41.1 | No |
| Propanal | -39.2 | -39.1 | -44.1 | -47.0 | -47.0 | -45.10±0.18 | Yes |
| Propanone | -47.2 | -47.8 | -47.0 | -55.0 | -55.0 | -52.2±0.1 | Yes |
| $CH_3(CC)_2NC$ | 158.4 | 158.4 | 159.1 | 150.6 | 150.6 | | No |
| $CH_3(CC)_2CN$ | 131.9 | 131.1 | 131.7 | 125.3 | 125.3 | | Yes |

Table 10: $\Delta_f H^O$ for isomers with 11 atoms and current astronomical status

| Molecule | Enthalpy of formation | | | | | | Astronomical status |
|---|---|---|---|---|---|---|---|
| | W1U | W2U | G3 | G4MP2 | G4 | Expt | |
| Dimethyldioxirane | -16.4 | -16.4 | -26.1 | -27.7 | -27.7 | | No |
| Glycidol | -48.3 | -48.3 | -58.5 | -57.3 | -57.3 | | No |
| Dioxolane | -61.6 | 62.1 | -71.2 | -73.3 | -73.3 | -72.1±0.5 | No |
| Lactaldehyde | -72.9 | -72.9 | -82.0 | -81.3 | -81.3 | | No |
| Methyl acetate | -84.8 | -84.7 | -91.7 | -95.1 | -95.1 | -98.0[97] | Yes |
| Ethylformate | -87.4 | -87.4 | -94.4 | -97.5 | -97.5 | -95.1 | Yes |
| *Propanoic acid* | *-100.6* | *-100.5* | ***-108.7*** | *-109.4* | *-109.4* | *-108.9±0.5* | *No* |

Table 11: $\Delta_fH^O$ for isomers with 12 atoms and current astronomical status

| Molecule | Enthalpy of formation | | | | | | Astronomical status |
|---|---|---|---|---|---|---|---|
| | W1U | W2U | G3 | G4MP2 | G4 | Expt | |
| *Ethyl methyl ether* | -47.2 | -47.1 | **-52.8** | -57.4 | -57.4 | *-51.9±0.2[88]* | Yes |
| Propanol | -53.7 | -53.4 | -61.3 | -61.9 | -61.9 | -60.2±0.7 | No |
| Propan-2-ol | -57.2 | -57.2 | -65.3 | 65.6 | -65.6 | -65.2 | No |
| 2-azabicyclo(2.1.0)pentane | 67.4 | 67.4 | 59.8 | 57.1 | 57.1 | | No |
| N-methylpropargylamine | 61.6 | 61.6 | 57.2 | 54.5 | 54.5 | | No |
| 3-butyn-1-amine | 54.2 | 54.2 | 60.8 | 48.4 | 48.4 | | No |
| N-methyl-1-propyn-1-amine | 54.4 | 53.9 | 53.0 | 46.6 | 46.6 | | No |
| N-vinylazaridine | 54.8 | 54.9 | 51.1 | 45.9 | 45.9 | | No |
| 2,3-butadiene-1-amine | 49.7 | 49.7 | 47.8 | 43.3 | 43.3 | | No |
| But-1-en-1-imine | 35.6 | 35.7 | 34.6 | 27.8 | 27.8 | | No |
| 2,2-dimethylethylenimine | 34.1 | 34.2 | 38.4 | 25.7 | 25.7 | | No |
| 3-pyrroline | 34.5 | 34.6 | 28.5 | 25.4 | 25.4 | | No |
| 2-aminobutadiene | 30.3 | 30.4 | 27.6 | 24.4 | 24.4 | | No |
| 2-isocyanopropane | 33.2 | 33.5 | 21.7 | 24.4 | 24.4 | | No |
| Propyl cyanide | 13.8 | 13.8 | 7.7 | 5.6 | 5.6 | 7.4 | Yes |
| Isopropyl cyanide | 13.6 | 13.6 | 6.7 | 5.2 | 5.2 | 5.4[97] | Yes |

**Figures**

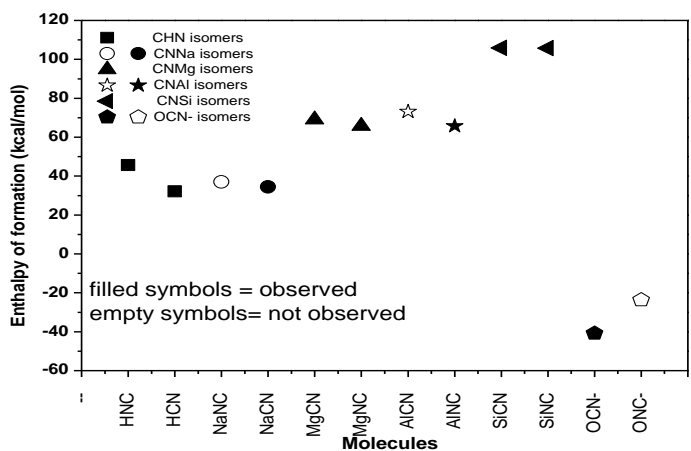

Figure 1: Plot showing the $\Delta_f H^O$ for molecules with 3 atoms

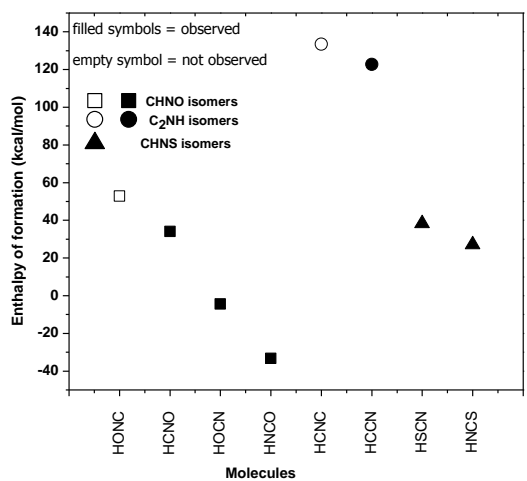

Figure 2: Plot showing the $\Delta_f H^O$ for molecules with 4 atoms

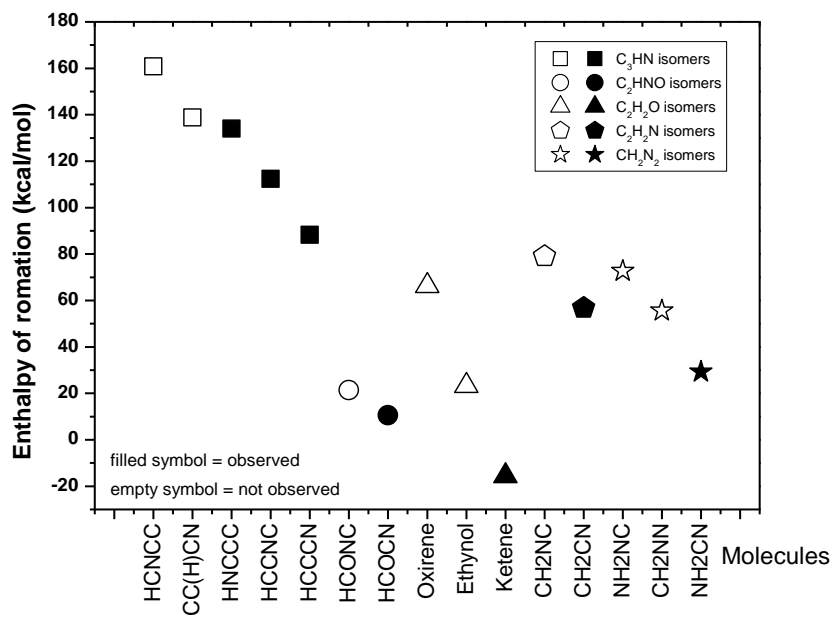

Figure 3: Plot showing the $\Delta_f H^O$ for molecules with 5 atoms

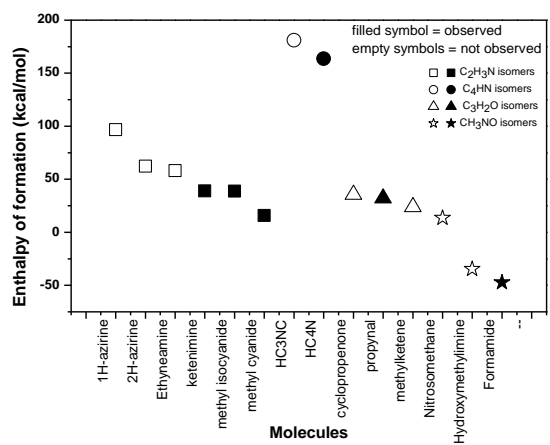

Figure 4: Plot showing the $\Delta_f H^O$ for molecules with 6 atoms

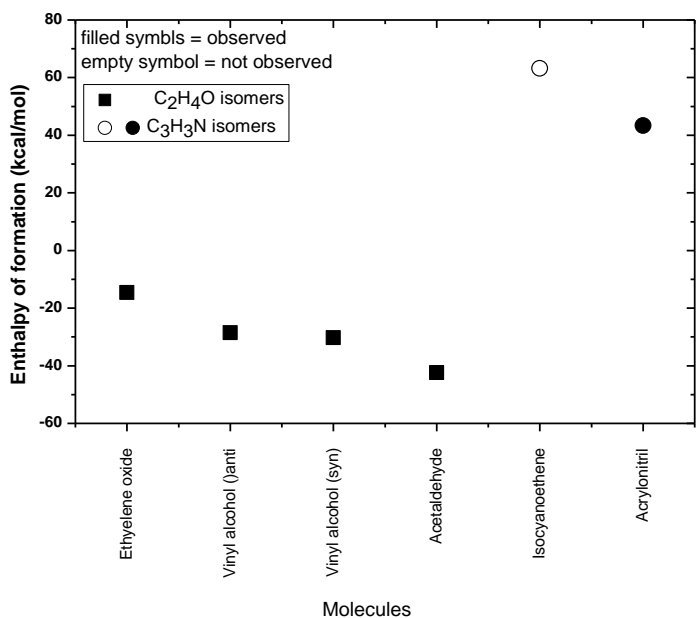

Figure 5: Plot showing the $\Delta_f H^O$ for molecules with 7 atoms.

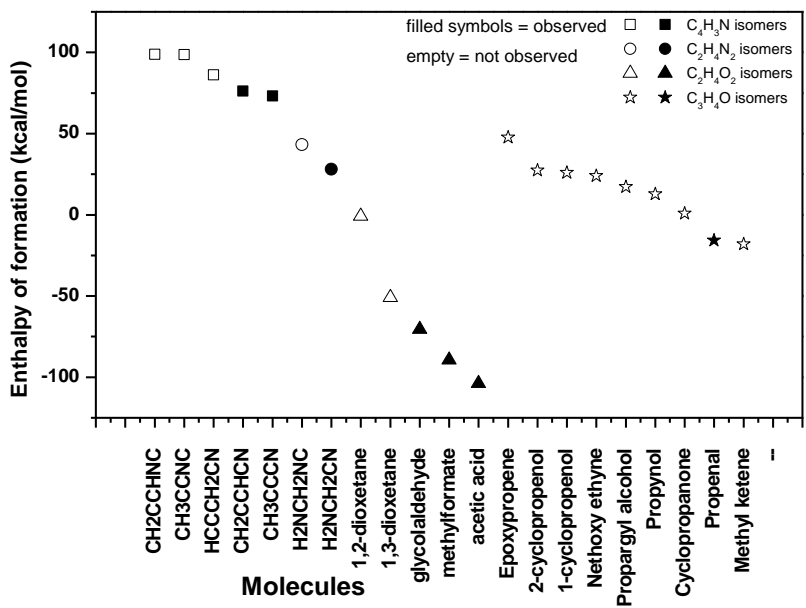

Figure 6: Plot showing the $\Delta_f H^O$ for molecules with 8 atoms

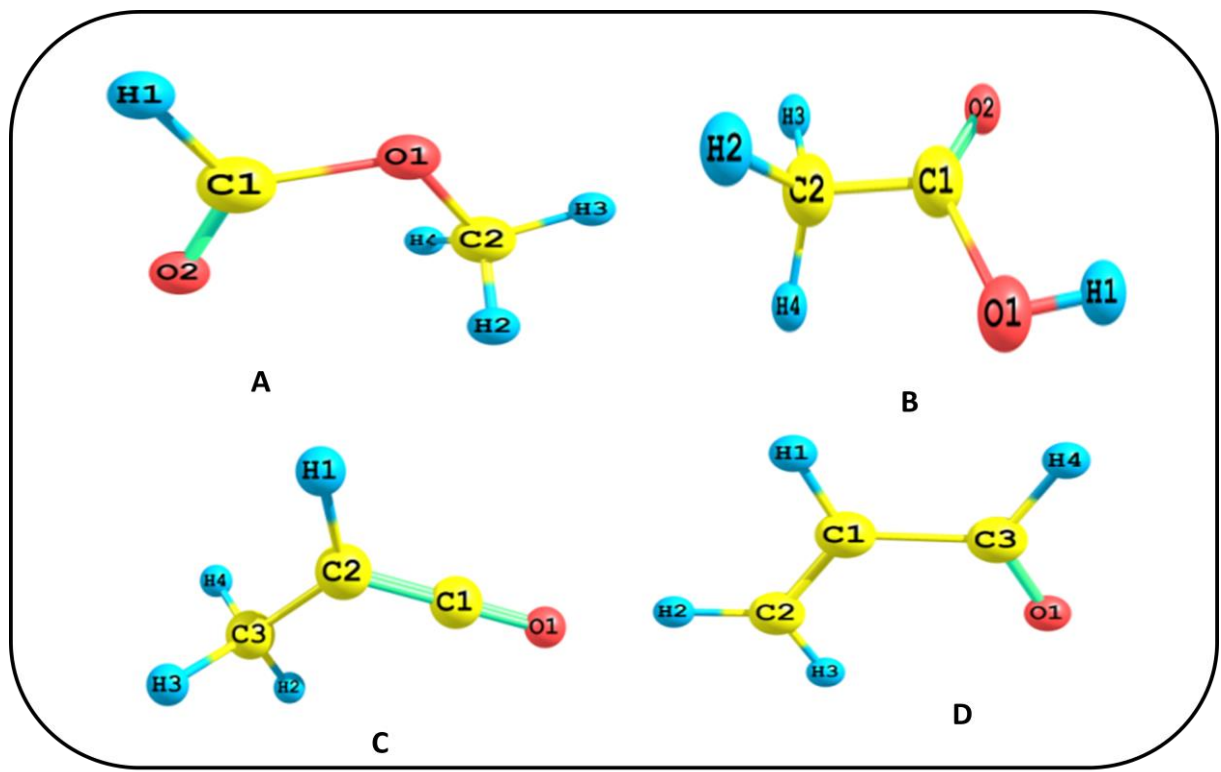

Figure 7: Optimized structures of methyl formate (A), acetic acid (B), methyl ketene (C) and propenal (D) at G4 level of theory.

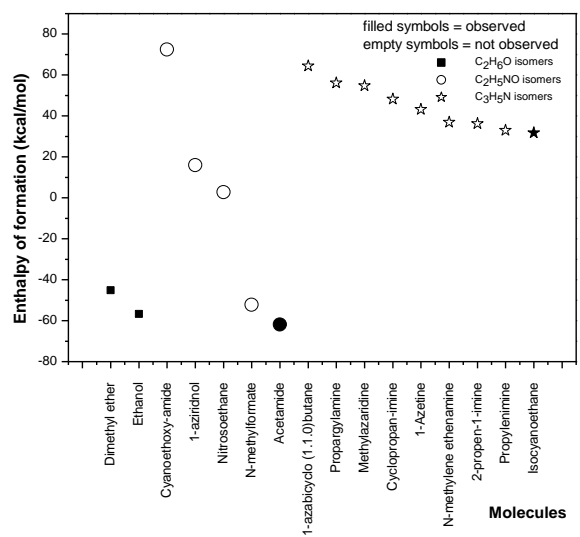

Figure 8: Plot showing the $\Delta_f H^O$ for molecules with 9 atoms

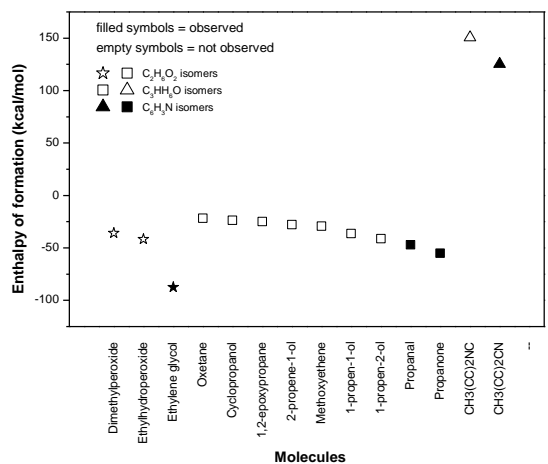

Figure 9: Plot showing the $\Delta_f H^O$ for molecules with 10 atoms

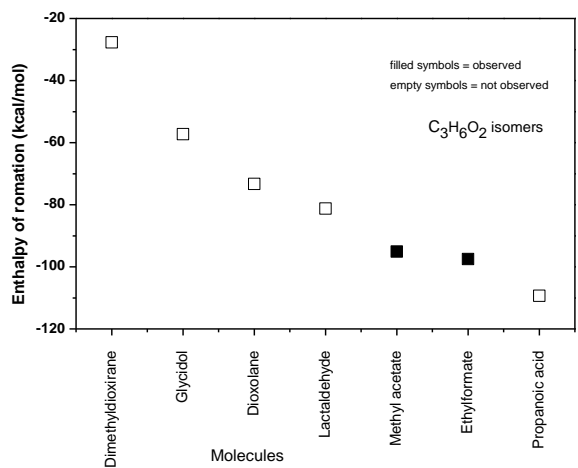

Figure 10: Plot showing the $\Delta_f H^O$ for molecules with 11 atoms

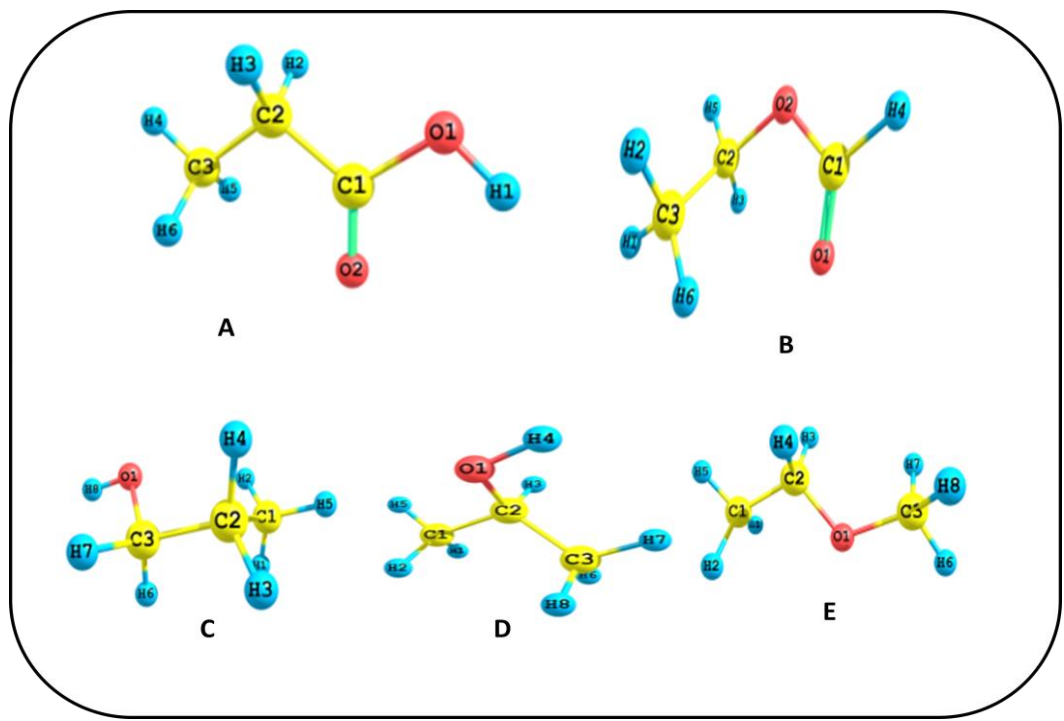

Figure 11: Optimized structures of propanoic acid (A), ethylformate (B), propanol (C), propan-2-ol (D) and ethyl methyl ether (E) at G4 level of theory.

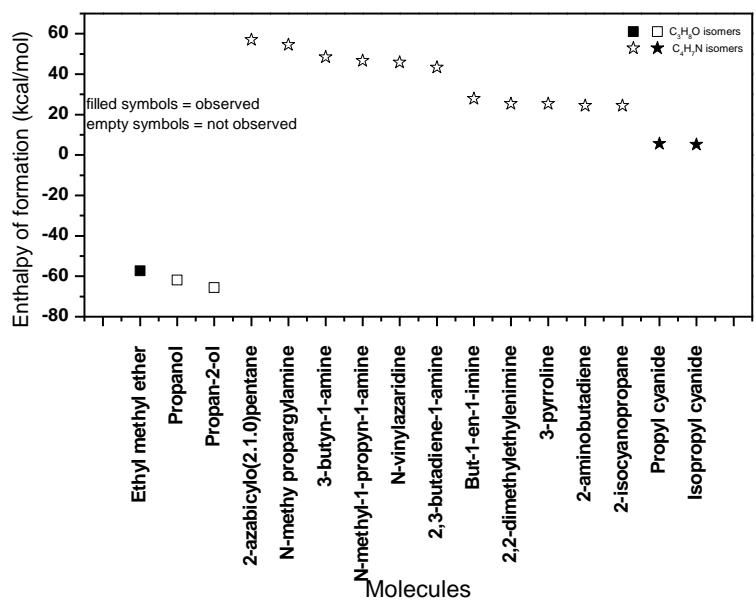

Figure 12: Plot showing the $\Delta_f H^O$ for molecules with 12 atoms